\documentclass{ws-mpla}

\usepackage{graphicx}
\usepackage{fancybox}
\usepackage{hyperref}

\newcommand{\MOM}{\mathrm{MOM}}

\newcommand{\alphaMOM}{\alpha_s^{\MOM}}

\newcommand{\ratio}{\mbox{$\frac{\langle\overline{s}s\rangle}{\langle\overline{u}u\rangle}$}}

\newcommand{\gluecond}{\mbox{$a^4 \langle \alpha_s G^2 / \pi \rangle  $}}

\begin{document}

\catchline{}{}{}{}{}

\title{Strong coupling and quark masses from lattice QCD}
\author{\footnotesize Craig McNeile}

\address{
Bergische Universit\"at Wuppertal, \\
Gaussstr.\,20, D-42119\\
Wuppertal, Germany.\\
mcneile@uni-wuppertal.de}

\maketitle

\pub{Received (Day Month Year)}{Revised (Day Month Year)}

\begin{abstract}

I review lattice QCD calculations of the strong
coupling and quark masses.

\end{abstract}

\ccode{PACS Nos.: 12.38.Gc; 12.15.Ff.}



\section{Introduction} \label{se:intro}

I review recent lattice QCD calculations
of the strong coupling and quark masses.
It is very timely to do this because the errors on these quantities
from the PDG~\cite{Beringer:1900zz} have been
reduced this year.

The basic theory behind the calculation of 
the properties of bound states of QCD, using lattice QCD,
is the creation and destruction of particles 
in the path integral formalism.
\begin{equation}
c_{ij}(t)=\frac{1}{Z}\int du\int d\psi\overline{d\psi}O(t)_{i}O(0)_{j}^{\dagger}e^{-S_{F}-S_{G}}
\label{eq:pathint}
\end{equation} 
where the $S_F$ is the action for the quarks and
$S_G$ is the action for the gauge fields.
The path integral is regulated by the introduction of a
space-time lattice and 
is computed in 
Euclidean space
using Monte Carlo techniques on the computer.
The final stage of the calculation is a set of 
physical quantities, such as a meson mass, for
a specific lattice coupling, volume and input quark
masses. The physical quantities are extrapolated 
to the continuum limit, infinite volume and 
physical quark masses.

\begin{table}[tb]
\caption{Parameters of two recent lattice QCD calculations.\label{tb:params}}
\centering
\begin{tabular}{|c|c|c|c|c|c|c|} \hline
Action       &  $n_f$  & \multicolumn{2}{|c|}{lattice spacing} &
\multicolumn{2}{|c|}{$m_{\pi}$}  & $ m_\pi L$ \\ \cline{3-6}
             &         & No. &  Range fm     & No. & Range MeV &  \\ 
\hline
HISQ~\cite{Bazavov:2012xda}   &  2+1+1  & 3 & 0.15 - 0.06 & $>3$   & 128
- 312 & 3.22 - 5.36
 \\
2 HEX Clover~\cite{Durr:2010aw} & 2+1 & 5 & 0.12 - 0.065 & $>5$ &
120 - 670 & 3.0 - 6.36  \\
\hline
\end{tabular}
\end{table}

The solution of Eq.~\ref{eq:pathint} solves QCD, at the 
input parameters, however the key issue is how physical the parameters
are and so for example whether a controlled continuum limit can be taken.
The computational cost of the 
lattice QCD calculation depends 
particularly on the masses
of the light quarks and
on the type of lattice approximation to the Dirac operator
used. 
In table~\ref{tb:params} I report the parameters
of two recent lattice QCD calculations
(see ~\cite{Hoelbling:2011kk} 
for a more complete review). 

It is important to 
compute experimental numbers as a cross-check on the lattice
QCD calculation.
BMW-c~\cite{Durr:2008zz} have accurately computed
10 ground state hadron masses from a 2+1
unquenched lattice QCD calculation.
The continuum limit was taken and the 
lightest
pion mass was 190 MeV. 
The HPQCD collaboration~\cite{Gregory:2009hq} 
have published a summary
plot of the masses of 22 mesons, some of which are predictions, which include the 
bottom or charm quarks.

I exclusively focus on the results of recent lattice QCD calculations.
Motivated by the FLAG~\cite{Colangelo:2010et}
group's review system,
I plot the lattice
results for quark masses in green if the systematics are under reasonable
control, but use red if more work is required, because for 
example only one lattice spacing is used. For the $\alpha_s$ summary
I use green or red circles for the quality of the continuum and 
chiral limit.

\section{Quark masses from lattice QCD}

Quark masses are inputted to the lattice QCD calculation
and a subset of the hadron spectrum is computed in 
lattice units. The quark masses are chosen to
reproduce the masses of a few hadrons. The lattice spacing
is also determined from a physical quantity.
After the bare quark masses have been determined 
they need to be converted into a continuum scheme.
\begin{equation}
m_q^{\overline{MS}}(\mu) = Z^{\overline{MS}}_m(\mu) m_q^{bare}
\label{eq:ZZ}
\end{equation}
The $Z^{\overline{MS}}_m(\mu)$ factors in Eq.~\ref{eq:ZZ} can be computed 
using lattice perturbation theory, or a variety of 
numerical techniques~\cite{Martinelli:1994ty,Jansen:1995ck,Lee:2013mla},
which compute the lattice contribution to $Z^{\overline{MS}}_m(\mu)$
to all orders. The quark mass is usually required at a standard
scale, such as 2 or 3 GeV, or the mass of the quark itself. 
The numerical technique to compute $Z^{\overline{MS}}_m(\mu)$
usually only works for a window of momentum, so some 
evolution of the quark masses with scale may be required.

The ratios of quark masses, such as 
$m_c/m_s$, $m_b/m_c$, $2 m_s/(m_d+m_u)$ are independent of
renormalisation, hence are now commonly computed.



\section{The QCD coupling}

As in the continuum the basic idea of determining
the QCD coupling is to compute some quantity $\sigma$
in the lattice QCD calculation, which has a perturbative
expansion.
\begin{equation}
\sigma =  a_0 + \alpha_s a_1 + \alpha_s^2 a_2 + \alpha_s^3 a_3 + ....
\end{equation}
The coefficients $a_i$ are computed in perturbation theory
but also typically contain non-perturbative contributions. 
Lattice perturbation theory can be used, or it is 
usually easier to take the continuum limit of a quantity and 
then use continuum perturbation theory, because higher
order calculations are easier in the continuum.

The contribution of the charm and bottom quarks
are perturbatively added to
$\alpha_s$ extracted from
a lattice QCD calculation with 2+1 flavors
of sea quarks. The coupling is then evolved 
typically using perturbation theory to the mass of the
Z boson. The inclusion of a quark's contribution using
perturbation theory is done at the scale of quark mass.
There are concerns that perturbation is not reliable 
at the scale of the charm mass (1.2 GeV), although 
this matching is not a significant contribution to
the error from~\cite{Davies:2008sw}. The matching
of the results for $\alpha_s$ from calculations
with $n_f=2$ to $n_f=5$ is more problematic 
(see Ref.~\cite{Sternbeck:2012qs} for a review plot which
includes $n_f=2$ results),
because it must be done at the scale of 100 MeV.

The HPQCD collaboration~\cite{Davies:2008sw}
have computed $\alpha_s$ by
measuring 22 different
combinations of small
Wilson loops in the numerical lattice
QCD calculation and comparing them to 
expressions in lattice perturbation theory.
The perturbative calculation included terms
of $\alpha_s^3$ and
has been
checked by Trottier and Wong using a numerical technique~\cite{Wong:2005jx}.
There are small non-perturbative contributions to 
Wilson loops from condensates, such as 
the non-perturbative gluon condensate,
which increase with the 
size of the loop.
The lattice spacing was obtained from the $\Upsilon$ spectrum and
$f_\pi$.
The calculation used data at 6 different lattice spacings,



The HPQCD collaboration~\cite{Allison:2008xk,McNeile:2010ji} 
have used the moments of correlators 
to compute the masses of the charm and 
bottom quarks, as well as the strong coupling.
In the lattice QCD calculation the time moments
of the correlator in Eq.~\ref{eq:pathint} are measured.
\begin{equation}
   G_n \equiv \sum_t (t/a)^n c_{11}(t)
\end{equation}
The moments of the correlators are analyzed
using continuum perturbation theory after the 
continuum limit is taken. The $G_4$ 
moment is insensitive to the heavy quark mass, so 
it is used to estimate $\alpha_s$.
The first paper~\cite{Allison:2008xk} on using moments
in an unquenched calculation
was a collaboration between the 
HPQCD collaboration and Chetyrkin, K\"{u}hn, Steinhauser,Sturm.
There was an updated result~\cite{McNeile:2010ji}.


The JLQCD collaboration have extracted $\alpha_s$
from the light vacuum polarization~\cite{Shintani:2010ph}.
\begin{equation}
  \langle J^a_\mu J^b_\nu\rangle(Q) = \delta^{ab}
  \left[
    (\delta_{\mu\nu}Q^2-Q_\mu Q_\nu)\Pi_J^{(1)}(Q) 
    - Q_\mu Q_\nu \Pi_J^{(0)}(Q) 
  \right],
\end{equation}
and $Q_\mu$ is space-like and
$J_\mu=V_\mu$ vector and $J_\mu=A_\mu$
for the axial current.
The correlators for the sum $\Pi_V(Q)+\Pi_A(Q)$ were fitted to an OPE 
based formulae using
continuum perturbation theory known up the 4 loops
for some cases. The calculation also used condensates,
either measured from other lattice calculations or 
taken from phenomenology. The continuum limit was not taken.


The strong coupling has 
recently been computed by Bazavov et al.~\cite{Bazavov:2012ka}
from the measured static QCD potential on the
lattice.
The static energy has recently been computed to $N^3LL$ 
in perturbation theory~\cite{Smirnov:2009fh,Anzai:2009tm}.
The basis of the lattice QCD calculation was gauge
configurations generated by the 
HOT collaboration generated
      as the T=0 part of thermodynamic project.
Eight lattice spacings in the range: 
0.805 GeV $<$  $a^{-1}$ $<$ 2.947 GeV. 
were used.


Although lattice QCD is a gauge invariant method, it is 
possible to fix to a specific gauge, such as Landau gauge, and then to
compute quark, gluon or ghost propagators.
The ETM collaboration~\cite{Blossier:2012ef} and Sternbeck
et al.~\cite{Sternbeck:2012qs} are using this technique to estimate 
$\alpha_s$.
A Taylor like coupling~\cite{Sternbeck:2007br} is defined
from the measured $Z$ and $J$ factors
\begin{equation}
 \alphaMOM(q^2) = \alphaMOM(\mu^2)\, Z(q^2,\mu^2)\, J^2(q^2,\mu^2)
\label{eq:couplingCON}
\end{equation}
$Z$ and $J$ are defined from the 
gluon and ghost propagator propagators
\begin{eqnarray}
  D^{ab}_{\mu\nu}(q^2,\mu^2) & = &
 \delta^{abx}\left(\delta_{\mu\nu} - \frac{q_{\mu}q_{\nu}}{q^2}
 \right)\frac{Z(q^2,\mu^2)}{q^2}
\\
G^{ab}(q^2,\mu^2) & = &
 -\delta^{ab}\frac{J(q^2,\mu^2)}{q^2}
\end{eqnarray}

In principle the coupling in Eq.~\ref{eq:couplingCON} shows the
running
of the coupling with scale. However, for larger momentum there
are errors from the lattice spacing. For smaller 
$q^2$ there are possible non-perturbative contributions.
So the coupling can only be extracted in a window of momentum.
Sternbeck
et al.~\cite{Sternbeck:2012qs} are investigating
reducing the lattice spacing errors
at larger scale by using lattice perturbation theory.

The ETM collaboration~\cite{Blossier:2012ef} 
use another approach to
extract $\alpha_s$.
To extend the in fit region in $p$ where they can
fit from $5.5 < p < 6.8$ GeV to
$1.7 < p <  6.8$ GeV, they
introduce an additional fit term~\cite{Blossier:2012ef}
outside the OPE.
\begin{equation}
\alpha_T^{d}(p^2)  =  \alpha_T(p^2)  +  \frac{d}{p^6} 
\label{eq:glueadd}
\end{equation}
They tested this method in quenched QCD, but the 
modification in Eq.~\ref{eq:glueadd} is empirical.
This calculation includes the dynamics of 2+1+1 flavors
of sea quarks.


The ALPHA collaboration developed an elegant method
      to compute $\alpha_s$ using 
the Scr\"{o}dinger functional (SF) on the lattice.
The ALPHA collaboration have used SF method for $n_f$ = 0, and 2
sea quarks.
      Calculations with $n_f$ = 4 are in 
progress~\cite{Tekin:2010mm}.
The method has been used by 
PACS-CS collaboration~\cite{Aoki:2009tf}
to compute $\alpha_s$ from 2+1 
flavors of sea quarks.
The SF coupling is defined via a derivative of a boundary field.
The lattice QCD calculation computes the step scaling function
to evolve the coupling to double the scale.
The step scaling function evolves the coupling up to high scale 
where the conversion from the
SF scheme to $\overline{MS}$
is done with perturbation theory including $\alpha_s^3$.
The dominant error on the final result for
$\alpha_s(M_Z^2)$ was from two different continuum 
extrapolations: constant or linear in the lattice
spacing. The lowest pion mass used in the calculation was 
500 MeV. This is heavy by modern standards.
The mass of the $\Omega$
baryon was used to set the lattice spacing.

\begin{table}[tb]
\caption{Summary of lattice methods to compute $\alpha_s$}
\centering
\begin{tabular}{|c|c|c|c|c|} \hline
Method           & scale & range GeV & pert. & non-perturb.   \\ \hline
Wilson loops~\cite{Davies:2008sw}     & $\Upsilon$  $f_{\pi}$  & 2.1 - 14.7   & $\alpha_s^3$ & \gluecond \\
Charm moments~\cite{McNeile:2010ji}    & $\Upsilon$  $f_{\pi}$      &    3        &  $\alpha_s^3$  & \gluecond \\
Light vaccum pol~\cite{Shintani:2010ph} & $r_0$  $f_{\pi}$  $\Omega$      &  1.8        &  $\alpha_s^3$  & $\overline{\psi}\psi$ ,
\ratio,\gluecond \\
Static energy~\cite{Bazavov:2012ka}        & $r_1$ $f_\pi$   & 0.8 - 2.9         & $\alpha_s^4$  &
renormalons  \\
Schr\"{o}dinger funct.~\cite{Aoki:2009tf} &  $\Omega$  & 16 &  $\alpha_s^3$ & - \\
Glue/ghost~\cite{Blossier:2012ef,Sternbeck:2012qs} &  $f_\pi$  & 1.7 - 6.8 &  $\alpha_s^3$ & $\frac{d}{p^6} $ \\
\hline
\end{tabular}
\label{tb:alphasSUM}
\end{table}

\begin{figure}
\centering
\includegraphics[scale=0.5,angle=0]{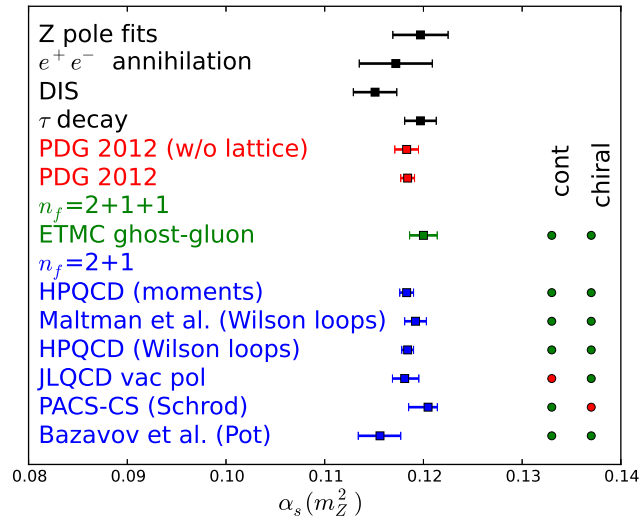}
\caption{Summary of $\alpha_s$ measurements from lattice QCD and 
summaries from other methods.}
\label{fg:summary}
\end{figure}

In table~\ref{tb:alphasSUM} I summarize the different
methods of extracting $\alpha_s$ from lattice QCD
and the summary of results are in figure~\ref{fg:summary}.
The color coding is different to the quark mass summary
plots and is explained in section~\ref{se:intro}.
I also include the 
PDG summary values~\cite{Beringer:1900zz} for different non-lattice methods for determining
$\alpha_s$.

The final summary for $\alpha_s(M_Z^2)$ 
from the PDG~\cite{Beringer:1900zz} is 
0.1184(7). If the lattice results were not included
in the average the result $\alpha_s(M_Z^2)$ = 0.1183(12) was obtained, 
so the 
lattice results for $\alpha_s$ are consistent with 
other methods, but they are important for the final
error. One of the goals of running a linear collider at
the Z peak (the GigaZ option~\cite{Allanach:2004ud}) 
was to measure 
$\alpha_s(M_Z^2)$ to an accuracy of 0.001, which
is already achieved if the PDG error is not
underestimated.

Many people worry that the error for $\alpha_s$ 
is dominated by the results from lattice QCD,
and one reviewer~\cite{Altarelli:2013bpa} 
of $\alpha_s$ measurements reported,
without any reasons, that he didn't believe
the error on the lattice results.
The lattice calculation using Wilson loop by 
Maltman et al.~\cite{Maltman:2008bx} was partly motivated to check
an earlier calculation of $\alpha_s$ by the HPQCD collaboration.
The mini-proceedings of a recent workshop~\cite{Bethke:2011tr}.
on $\alpha_s$ provide a useful summary of the views and
results of the experts.

From table~\ref{tb:alphasSUM}
the lowest scale used to determine $\alpha_s$
is 0.8 GeV. There have been speculations that the 
coupling will evolve to a 
constant for low 
scales (for example~\cite{Ermolaev:2012xi}).
There have been some lattice QCD calculations 
to investigate the coupling defined in 
Eq.~\ref{eq:couplingCON} for small momentum.
The motivation for these calculation is 
to test confinement models and different
solutions of Schwinger-Dyson equations.

Momentum on the lattice is quantized in units
of $p= \frac{2\pi}{L}$ for a box side $L$,
so large spatial volumes (for example 
($\mbox{16 fm})^4$ in~\cite{Bogolubsky:2009dc} )
are required to reach small momentum.
The results of Bogolubsky et al.~\cite{Bogolubsky:2009dc} 
(see figure 5 of ~\cite{Bogolubsky:2009dc}) shows
that the coupling defined in Eq.~\ref{eq:couplingCON} 
goes to zero for small momentum.
This corresponds to the ``decoupling solution'' of 
Schwinger-Dyson equation. The requirement for large volumes 
means that the lattice QCD calculations are done in
quenched QCD ($n_f$=0) with coarse lattice spacing with no 
continuum limit.

Currently there many 
lattice calculations which are studying the evolution
of coupling in QCD like theories with a large number of sea
quarks or quarks in a adjoint representation to help 
 BSM model builders
using strongly interacting theories
 (see~\cite{Giedt:2012it} for
a recent review).

\subsection{$\alpha_s$ and the unification of couplings}

The error on the value of $\alpha_s$ is the
biggest error on the the unification of
the three standard model gauge couplings. 
Although the inclusion of additional
degrees of freedom from SUSY theories improves the
unification of all three coupling at a single scale,
the agreement is not perfect.

Some groups have tried to use the unification of couplings
as a way to discriminate between different SUSY models.
For example King et al.~\cite{King:2007uj} 
find a more accurate unification
of couplings in their exceptional
super-symmetric standard model,
than with the MSSM. There is even work with the 
unification of couplings in technicolor 
models~\cite{Gudnason:2006mk}.
One of the goals of the linear collider running
as the GigaZ option~\cite{Allanach:2004ud}
(as a Z factory) was to 
measure $\alpha_s$ more
accurately to test unification.
A report~\cite{AbelleiraFernandez:2012cc}
for the proposed Large Hadron Electron Collider
estimates that it may be possible to calculate 
$\alpha_s$ to an accuracy of 0.11 \% from DIS
and mention testing coupling unification as motivation.
The status of the coupling unification of the couplings is reviewed
by Raby in the GUT review in the PDG~\cite{Beringer:1900zz}.

\begin{table}
\centering
\caption{$\beta$ functions for standard model and MSSM. 
$n_g$ generations and $n_h$ Higgs doublets.}
\begin{tabular}{|c|c|c|} \hline
$b_i$     &  Standard model & MSSM  \\ \hline
$b_3$     & 11 - $\frac{4}{3} n_g$  & 9 - 2 $n_g$  \\
$b_2$     & $\frac{22}{3}$ -  $\frac{4}{3} n_g$ - $\frac{1}{6} n_h$  & 6 - 2 $n_g$ - $\frac{1}{2} n_h$  \\
$b_1$     & - $\frac{4}{3} n_g$ - $\frac{1}{10} n_h$  & -2 $n_g$ -
$\frac{3}{10} n_h$  \\  \hline
\end{tabular}
\label{tb:betafunc}
\end{table}

As an illustrative example I use one loop 
evolution equations of the
QCD and electroweak couplings as a function
of the scale $Q$. 
I essentially follow the example in the text
book~\cite{Aitchison:2007fn}
(see also Peskin~\cite{Peskin:1997ez}.)
\begin{equation}
1/\alpha_i(Q)   =  1/\alpha_i(M_Z) + \frac{b_i}{2\pi} \log(Q/M_Z)
\end{equation}
The first two couplings are defined by
\begin{eqnarray}
\alpha_2^{-1} & = & \alpha_{em}^{-1} \sin^2( \theta_W ) \\
\alpha_1^{-1} & = & \frac{3}{5} \alpha_{2}^{-1} \cot^2( \theta_W ) 
\end{eqnarray}
where $\alpha_{em}$ is the QED coupling and 
$\theta_W$ is the weak mixing angle. The third coupling
($\alpha_3$) is the QCD coupling $\alpha_s$.
The explicit values for $b_1$, $b_2$, and $b_3$ 
are in table~\ref{tb:betafunc} in terms of the 
number of Higgs bosons and generations.

\begin{figure}
\begin{minipage}[b]{0.45\linewidth}
\centering
\includegraphics[%
  scale=0.35, 
  angle=0,keepaspectratio=true,
  origin=t,clip]{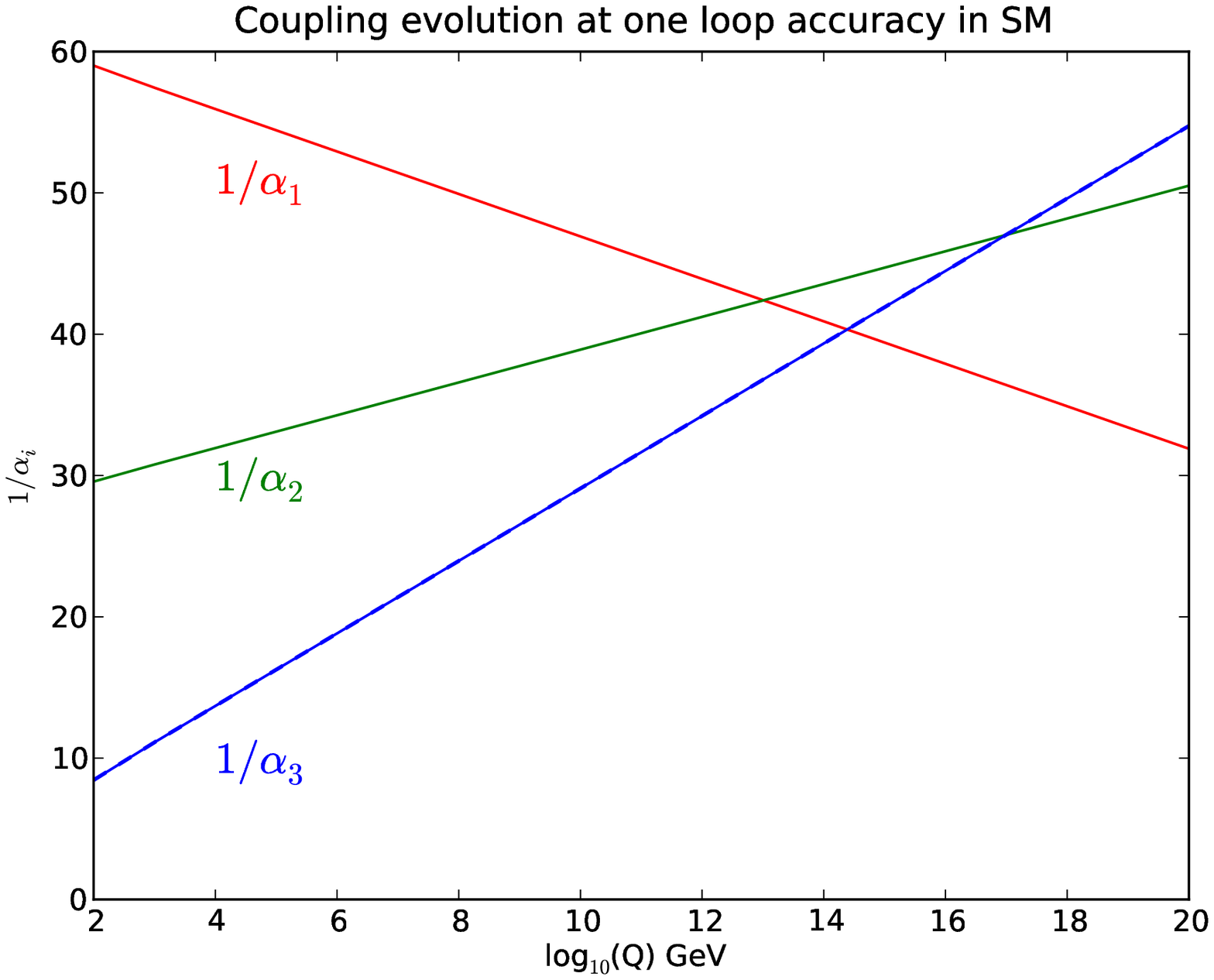}
\caption{Coupling unification in SM}
\label{fg:couplingSM}
\end{minipage}
\hspace{0.5cm}
\begin{minipage}[b]{0.45\linewidth}
\centering
\includegraphics[%
  scale=0.35, 
  angle=0,keepaspectratio=true,
  origin=t,clip]{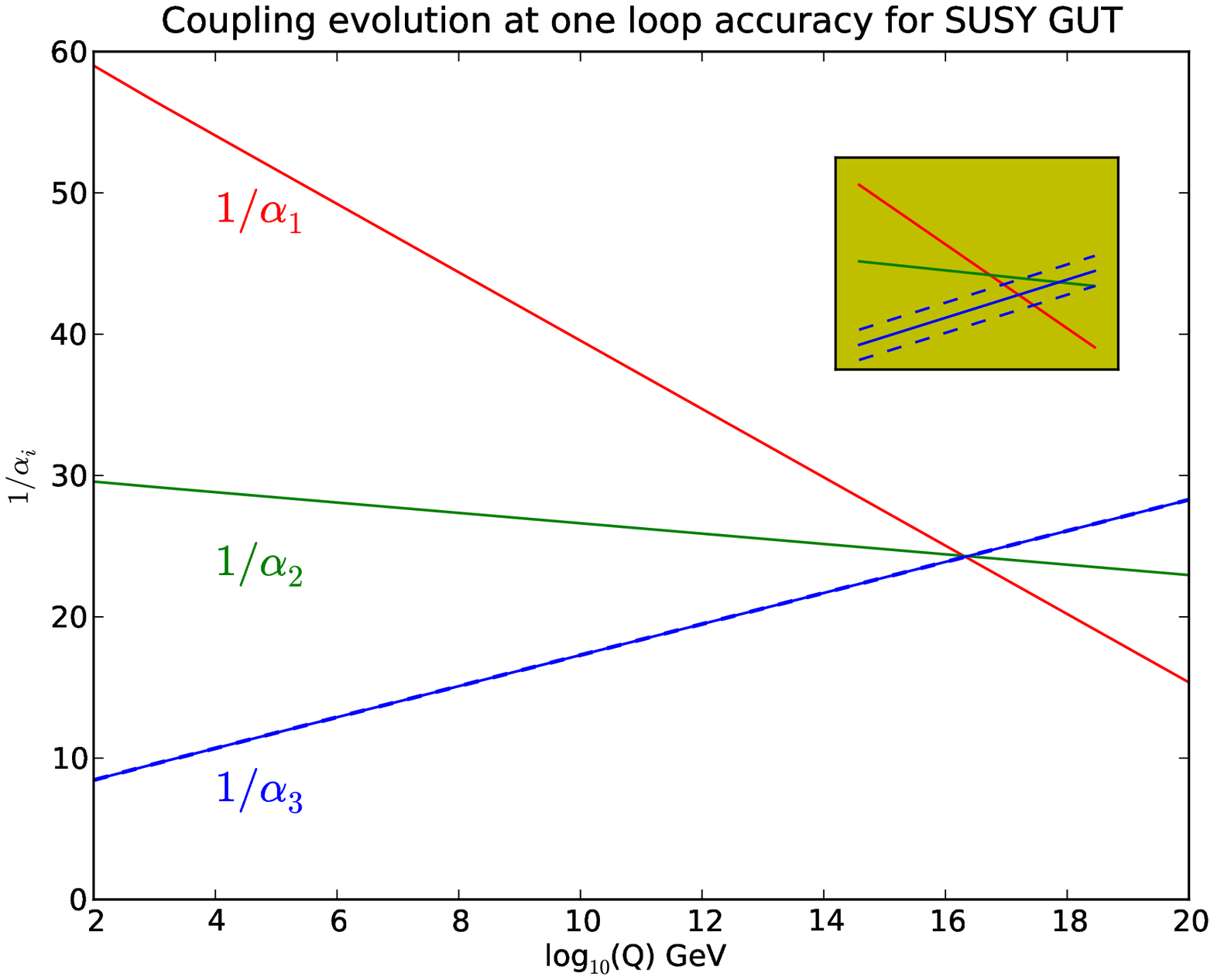}
\caption{Coupling unification in MSSM}
\label{fg:couplingMSSM}
\end{minipage}
\end{figure}

In figure~\ref{fg:couplingSM} I show the running
of the three couplings in the standard model and
in figure~\ref{fg:couplingMSSM}, I show the equivalent
plot for the MSSM, including the 1$\sigma$ error band
from $\alpha_s$ and the region around unification scale magnified.
It is clear that unification in the
MSSM is better than for the standard model, but that the
unification is not exact in the MSSM. I have not
investigated using two loop evolution equations, because
that involves threshold effects for new particles.
References to more detailed
calculations were discussed 
above~\cite{King:2007uj,Gudnason:2006mk,Allanach:2004ud}.

Once some evidence for BSM particles is found
and or proton decay is observed, then the error on $\alpha_s$
will be an important factor for tests of coupling unification,
and thus help to explore the physics at the unification scale.

The value of $\alpha_s$ is also important for
the stability of the Higgs potential~\cite{EliasMiro:2011aa}.

\section{The mass of the charm quark from lattice QCD} \label{se:charmmass}

\begin{figure}
\begin{minipage}[b]{0.45\linewidth}
\centering
\includegraphics[%
  scale=0.35, 
  angle=0,keepaspectratio=true,
  origin=t,clip]{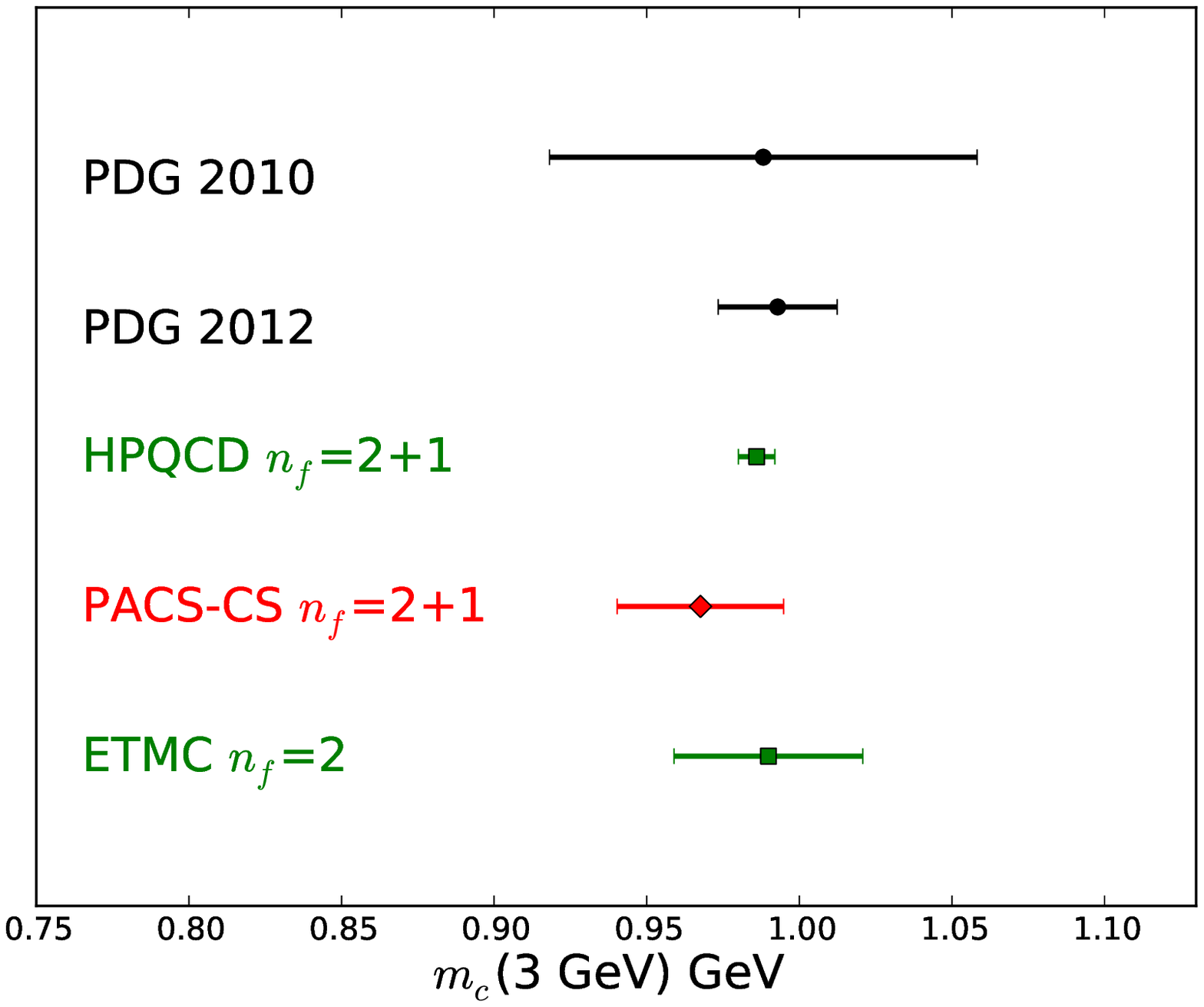}
\caption{Mass of the charm quark.}
\label{fg:charmsummary}
\end{minipage}
\hspace{0.5cm}
\begin{minipage}[b]{0.45\linewidth}
\centering
\includegraphics[%
  scale=0.35, 
  angle=0,keepaspectratio=true,
  origin=t,clip]{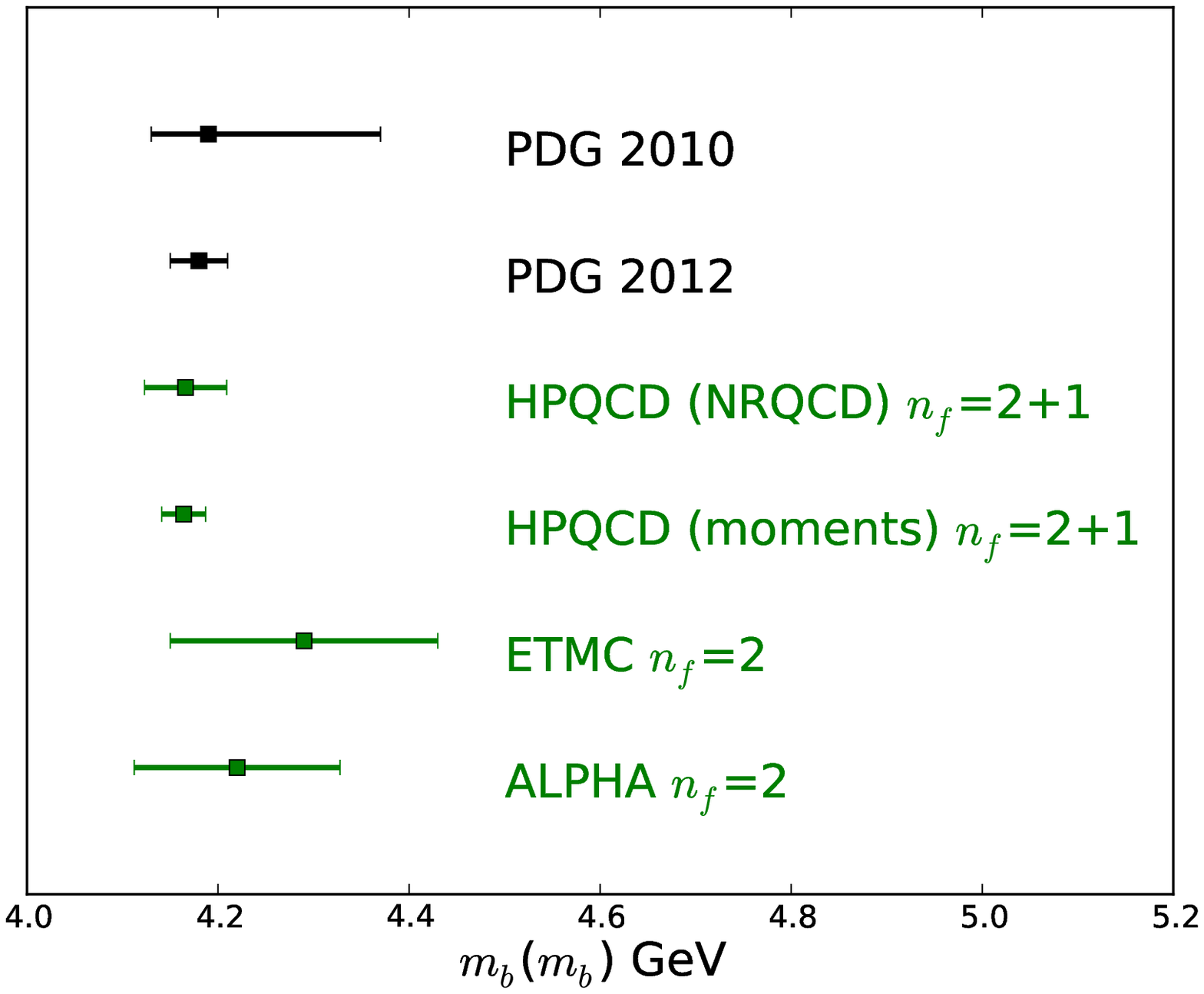}
\caption{Mass of the bottom quark.}
\label{fg:botsummary}
\end{minipage}
\end{figure}

The main complication for including the charm quark in
lattice QCD calculations is that the size of the 
charm mass in lattice units is not small, 
which can cause problems with the 
continuum extrapolation. Standard effective
field theories such as NRQCD or HQET 
are not useful for charm, but a 
lattice effective field theory developed by the 
FNAL group and others is commonly used.

I summarize
recent results for 
the mass of the charm quark in figure~\ref{fg:charmsummary}.
The HPQCD collaboration~\cite{McNeile:2010ji}
used the 
moments method to compute $m_c(\mbox{3 GeV})$.
The ETM collaboration~\cite{Blossier:2010cr} 
included 4 lattice spacings and
the renormalization of the quark mass was done 
with the Rome-Southampton technique.
The PACS-CS collaboration~\cite{Namekawa:2011wt}
have computed the
mass of the charm quark at single 
lattice spacing $a^{-1}=2.194(10)$ GeV, but using the
physical pion mass after re-weighting a lattice QCD
calculation with pion mass of 
of 152(6) MeV. The renormalisation was done 
by a non-perturbative method for the massless
factor and one loop perturbation theory for 
the massive renormalization contribution.

\section{The mass of the bottom quark from lattice QCD.}

Until very recently the mass of the bottom quark
was too big in lattice units for standard
lattice actions to be used, hence the majority 
of lattice QCD calculations used an effective field
theory, such as NRQCD or static QCD.
In figure~\ref{fg:botsummary} I show a summary plot for the mass of the
bottom quark from lattice QCD.

The HPQCD collaboration 
has recently published~\cite{Lee:2013mla} 
a result for $m_b(m_b)$ using Non-relativistic
QCD for the bottom quark. The 
$O(v^4)$ improved Symanzik action NRQCD action 
was used. The basic idea is to compute the 
binding energy of a heavy-light or heavy meson.
The mass of 
the bottom quark~\cite{Lee:2013mla}
in the $\overline{MS}$ scheme $m_b(\mu)^{\overline{MS}}$
is extracted from
\begin{equation}
m_b(\mu)^{\overline{MS}} = 
\frac{1}{2 Z_M(\mu)} \left( M^{expt}_{\Upsilon} - a^{-1}
( a E_{sim} - 2 a E_0 ) \right)
\label{eq:nrqcdmass}
\end{equation}
where $E_{sim}$ is the measured binding energy 
in the lattice calculation, 
$E_0$ is a lattice perturbative factor,
$M^{expt}_{\Upsilon}$ is the experimental mass of the Upsilon 
meson
and $Z_M(\mu)$ is the perturbative matching factor
between the pole mass and the $\overline{MS}$ 
mass in continuum perturbation theory.
The final perturbative expression for 
$m_b(\mu)^{\overline{MS}}$ should be free of
renormalon ambiguities, because of cancellation 
between the series for $Z_M(\mu)$ and $E_0$.
A value for $m_b$ was also extracted from the $B_s$ mass.

The lattice NRQCD action is complicated, so 
it is difficult to do higher order calculations of 
$E_0$ using lattice perturbation theory, instead
a mixed approach was taken.
The quenched diagrams
were done by doing the numerical calculation with very fine
lattice spacings. The $\alpha_s^2 n_f $ contributions were
obtained by
directly computing the 4 Feynman graphs in lattice
perturbation theory.
The numerical lattice QCD calculation used two
ensembles with lattice spacings: 0.09 and 0.12fm.

In the past the static (infinite mass) limit effective
field theory has been used to compute the mass of the
bottom quark.
The ALPHA collaboration developed an elegant method
to compute the $1/M_Q$ corrections to the static 
limit with a numerical method~\cite{Bernardoni:2012fd}.

The ETM collaboration~\cite{Blossier:2009hg,Dimopoulos:2011gx}
has developed a technique to 
take multiple ratios of the meson mass divided
by the quark pole mass (motivated by HQET).
This allowed
an extrapolation of their data up to the bottom quark mass.

The HPQCD collaboration has developed the 
relativistic 
HISQ action~\cite{Follana:2006rc}
      (with no $O(a^2)$ tree lattice spacings corrections)
The MILC collaboration~\cite{Bazavov:2009bb}
have generated gauge configurations
with the smallest lattice spacing of 0.045 fm.
These two developments allowed bare quark masses close to the physical
bottom mass to be used in the calculation with the relativistic action.
The moments method was used to compute
$m_b(m_b)$ in a similar manner to calculation
of the mass of the charm quark in section~\ref{se:charmmass}.
There was also a cross-check on the moments calculation
by taking ratios of $m_b / m_c$ from the quark mass
used in the action.

\section{A review of lattice reviews}

There are a number of groups which review lattice QCD
calculations of the masses of quark masses. There is 
a review section on quark masses in the PDG written 
by Sachrajda and Manohar. The FLAG group have provided
a comprehensive review of the light quark
masses~\cite{Colangelo:2010et}.  There is also
a group of three people called the 
Lattice averaging group~\cite{Laiho:2009eu}
(\url{http://latticeaverages.org/})
which provide averages of various quantities,
including the light quark masses.
FLAG 
joined by the lattice averaging group,
 are planning to
expand their review~\cite{Colangelo:2012iz}
to include the charm and bottom 
quark masses and $\alpha_s$.

In the past there have been reviews of quark masses
at either the annual lattice conference or other
conferences. For example Heitger~\cite{Heitger:2009qf} and
Davies~\cite{Davies:2013jt}. 

The quark masses are
usually quoted at standard scales, such as 2, 3 GeV or
the mass of the bottom quark, but these may be too small for BSM
model building. So there are summary tables of 
quark masses at 1 TeV and higher~\cite{Xing:2007fb},
which mostly use as input the quark masses 
from the PDG~\cite{Beringer:1900zz}.

One important issue is how to correctly average the lattice QCD
results. Most lattice results quote errors that are a mixture
of statistical or systematic errors.
The PDG~\cite{Beringer:1900zz} use 
a simple weighted mean without including any correlations
to average their data. The errors are increased if $\chi^2/dof > 1 $.
The are a number of possible causes 
for correlations between the results from different lattice QCD calculations.
For example, some lattice calculations use the same 
gauge configurations, but use different valence actions.
Sometimes an unphysical quantity such as $r_1$ calibrated by
another lattice QCD calculation is used by a number of
different calculations, which partially correlate the errors.
The lattice averaging group~\cite{Laiho:2009eu} 
have started to compute lattice QCD averages with
correlations (using the formalism in~\cite{Schmelling:1994pz}).
FLAG~\cite{Colangelo:2010et} argue that their error estimates 
are conservative enough to not require the use
of correlations.

A particularly important issue is what results to 
include. For example the lattice averaging group will
include the results presented at conferences, if in their 
judgement the error analysis is reliable, but FLAG only
include results published in journals.

\section{The mass of the strange and light quarks}

\begin{figure}
\begin{minipage}[b]{0.45\linewidth}
\centering
\includegraphics[%
  scale=0.35, 
  angle=0,keepaspectratio=true,
  origin=t,clip]{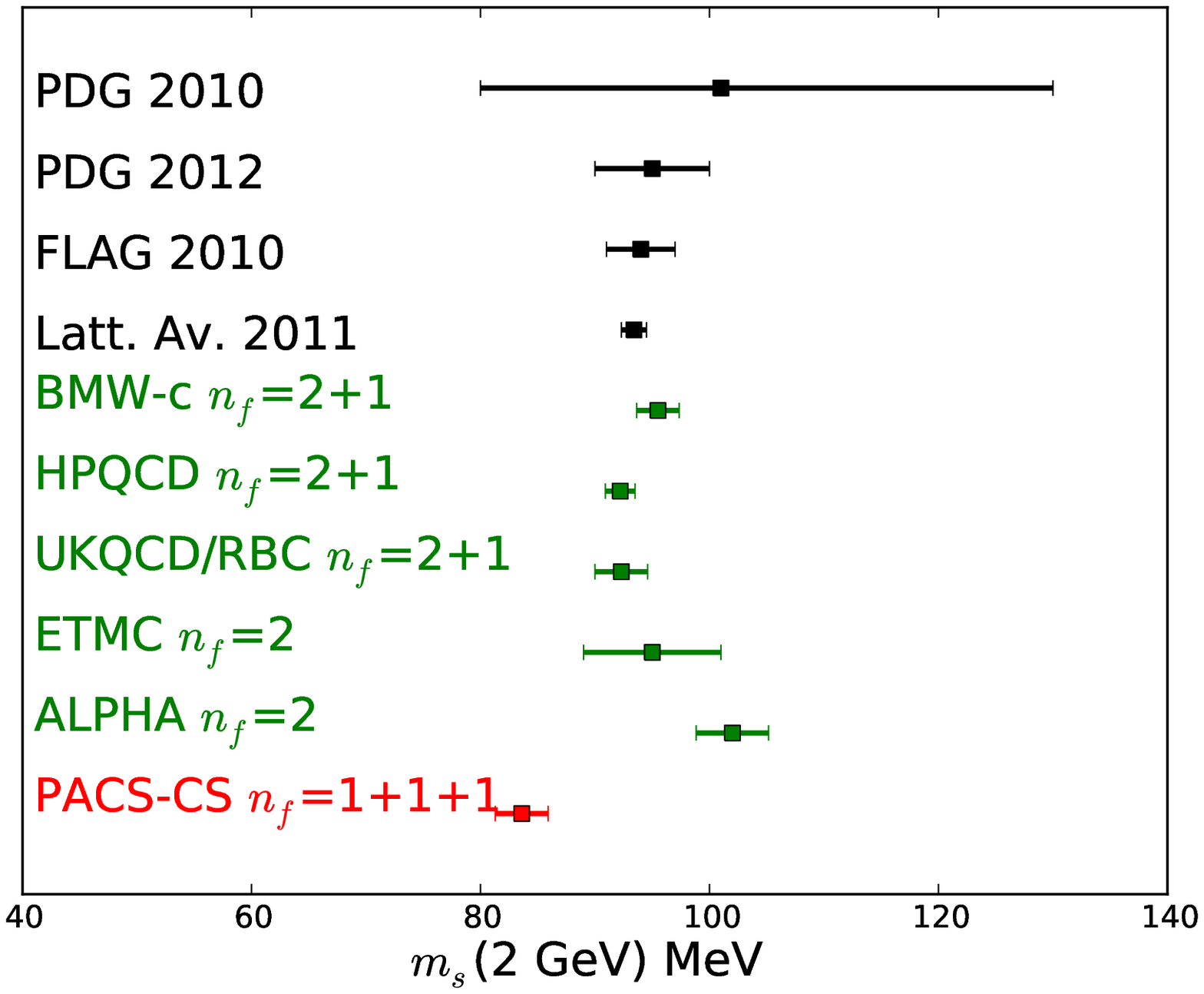}
\caption{Mass of the strange quark.}
\label{fg:smasssumm}
\end{minipage}
\hspace{0.5cm}
\begin{minipage}[b]{0.45\linewidth}
\centering
\includegraphics[%
  scale=0.35, 
  angle=0,keepaspectratio=true,
  origin=t,clip]{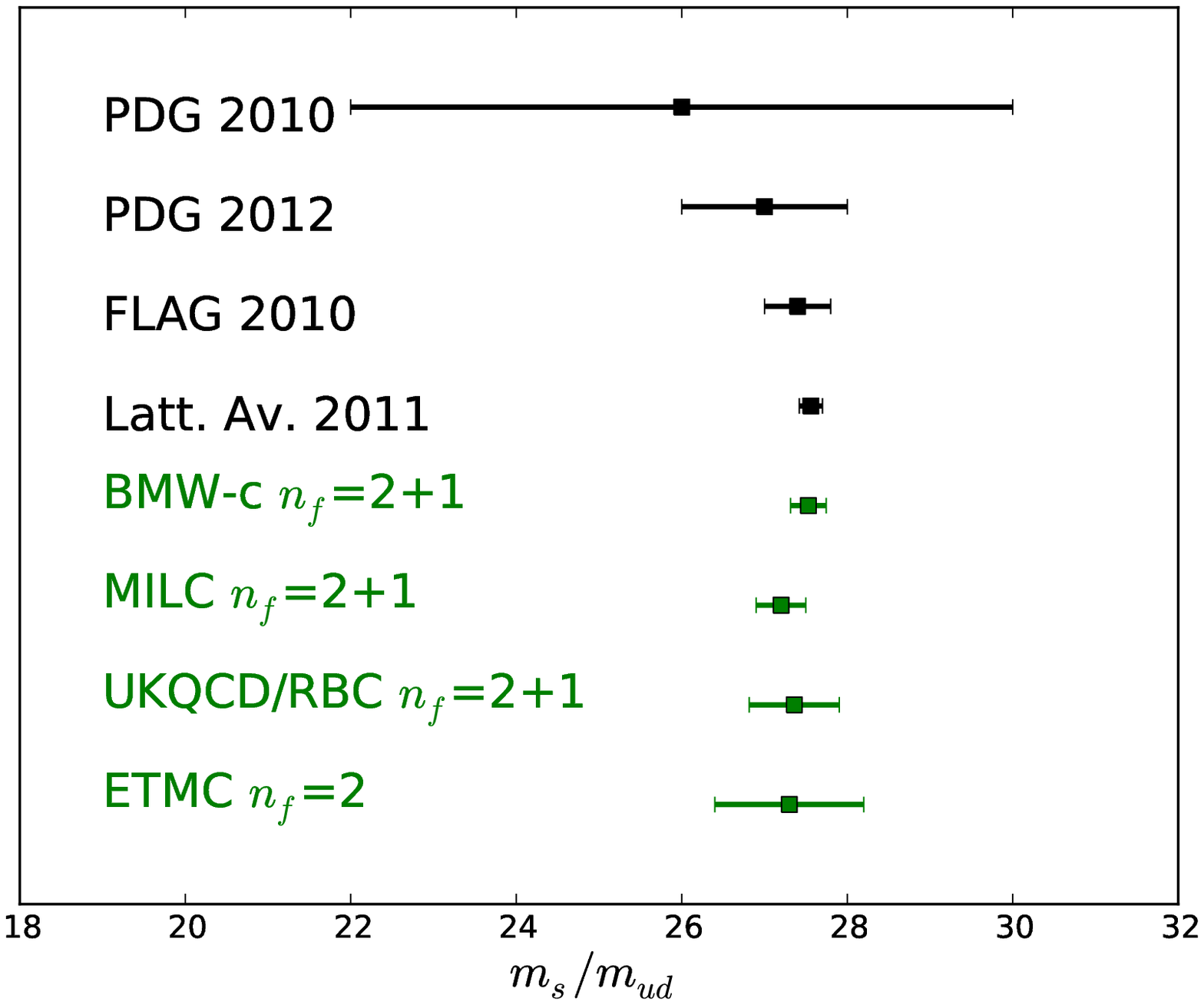}
\caption{Summary of $\frac{m_s}{m_{ud}}$}
\label{fg:msBYmlsumm}
\end{minipage}
\end{figure}

I plot a summary
of some recent lattice QCD calculations of
the mass of the strange quark in figure~\ref{fg:smasssumm} . 
I plot the ratio of the quark masses
$m_s$ to $m_{ud}$ (where $m_{ud} = \frac{m_u + m_d}{2}$)
from some recent lattice QCD calculations
in figure~\ref{fg:msBYmlsumm}.
The FLAG group~\cite{Colangelo:2010et} has
reviewed the determination of the masses of light quarks,
but does not include the more recent 
results~\cite{Aoki:2012st,Fritzsch:2012wq,Arthur:2012opa}.

BMW-c have computed the mass of the 
strange quark~\cite{Durr:2010vn}
from a lattice QCD calculation with 2+1 flavors of
sea quarks. The analysis included 5 lattice spacings, three
of which included ensembles at the physical pion mass. The Rome-Southampton
technique was used to compute the lattice part of the renormalisation
numerically~\cite{Martinelli:1994ty}. 
The UKQCD/RBC collaboration have 
recently reported~\cite{Arthur:2012opa}
the mass of the strange quark using a similar
technique to that used by BMW-c, but using heavier pion masses,
different lattice actions,
and only two lattice spacings. The ETM collaboration also
used the Rome-Southampton technique to renormalise their quark 
masses~\cite{Blossier:2010cr}.

The HPQCD collaboration~\cite{Davies:2009ih}
computed the mass of the strange quark by calculating 
the $m_c/m_s$ ratio in the continuum limit, and 
using $m_c$ computed 
with the moments 
method~\cite{Allison:2008xk,McNeile:2010ji}.

\section{Including QED and isospin violation in lattice calculations} \label{eq:qedisospin}

Currently the majority of lattice QCD
calculations only include $n_f$ = 2, 2+1 or
2+1+1 flavors of sea quarks. To extract the
individual up or down quark masses requires
QED and isospin violation 
from effective field theory
to be added to the results
of these calculations by hand 
(see~\cite{Colangelo:2012iz}
for a review).

There is a lot of development of
lattice 
calculations~\cite{Blum:2010ym,Ishikawa:2012ix,Portelli:2012pn,Aoki:2012st}
which directly include isospin and QED.
There are additional challenges to lattice QCD calculations
which include the dynamics of 1+1+1 sea quarks and 
the dynamics of QED.
For example including QED 
makes the calculations more noisy and
the renormalisation is more complicated.
The long range nature of QED may cause additional finite size
effects~\cite{Hayakawa:2008an}. The calculation
of the correlators for the $\pi^0$ require
the calculation of disconnected diagrams 
when $m_u \neq m_d$.

Including quenched QED in lattice QCD calculations
is done by multiplying the U(1) field, generated by
a separate lattice calculation, into the
QCD gauge fields.  
The RBC collaboration~\cite{Ishikawa:2012ix}
are using re-weighting to
move from a lattice QCD calculation with
2+1 sea quarks and no QED to one which includes the
QED dynamics in the sea. More ambitiously
the PACS-CS collaboration~\cite{Aoki:2012st}
are using re-weighting to move from a 2+1 
lattice QCD calculation to a 1+1+1+QED calculation.
The RM123 
collaboration~\cite{deDivitiis:2011eh,deDivitiis:2013xla}
have developed a formulation that
measures additional correlators, based on an 
expansion in $m_u - m_d$ and including 
quenched QED~\cite{deDivitiis:2013xla}.


\begin{figure}
\centering
\includegraphics[%
  scale=0.4, 
  angle=0,keepaspectratio=true,
  origin=t,clip]{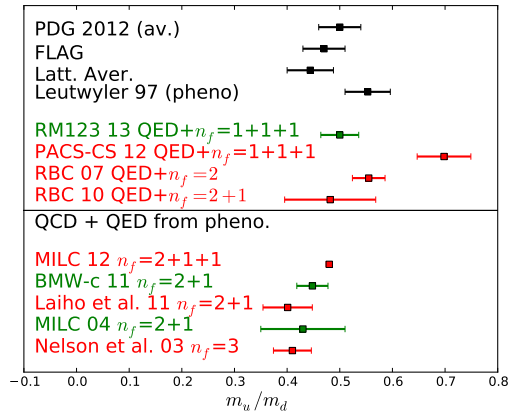}
\caption{Summary of recent values for $m_u/m_d$}
\label{fg:mumdsummary}
\end{figure}

I summarize in figure~\ref{fg:mumdsummary} 
the values of the ratio of $m_u/m_d$ from lattice QCD
and non-lattice methods included in the PDG summary of 
$m_u/m_d$. I include results from review 
groups~\cite{Colangelo:2012iz,Laiho:2009eu,Beringer:1900zz}, 
lattice calculations which include the dynamics of 
QED in 
someway~\cite{Blum:2007cy,Blum:2010ym,deDivitiis:2013xla,Aoki:2012st},
and lattice calculations which include
the QED 
from 
phenomenology~\cite{Bazavov:2012dg,Laiho:2011np,Durr:2010aw,Aubin:2004fs,Nelson:2003tb}.
The PACS-CS result for $m_u/m_d$ was from a calculation at only one lattice
spacing~\cite{Aoki:2012st}.
I am disappointed that the errors on $m_u/m_d$ have 
not been reduced by much since the result by MILC
in 2004~\cite{Aubin:2004fs}.


\subsection{Testing textures with quark masses}

Ideally I would like to understand 
why the CKM matrix is diagonally dominant and
the hierarchy of quark masses. 
Unfortunately, 
currently there are no compelling BSM models 
which make definite and falsifiable
predictions for the masses of the 
quarks. 
There has been a long history~\cite{Fritzsch:1977vd}
of looking for
patterns in the CKM matrix elements and quark
masses from textures.
This is numerology, but may give some hints.
There are reviews of 
textures~\cite{Fritzsch:1999ee,Babu:2009fd,Gupta:2013yha}.
I hope that the reduced errors on the 
masses of the quarks from lattice QCD calculations
will help constrain some of the proposed relations between CKM matrix
elements and the masses of the quarks.

As one example,
I use lattice QCD results to test one proposed connection
between the $V_{us}$ CKM matrix elements and 
the massess of the up and strange quarks,
from the model by Chkareuli and Froggatt~\cite{Chkareuli:1998sa}
(other relations from Ref.~\cite{Chkareuli:1998sa} were tested in
Ref.~\cite{McNeile:2010xe}.)
\begin{equation}
V_{us} \sim \sqrt{\frac{m_d}{m_s}} = \sqrt{\frac{2}{(1 + m_u/m_d) m_s / m_{ud}}  }
\label{eq:Vus}
\end{equation}
There are possible corrections to Eq.~\ref{eq:Vus} which
can be tested with more accurate values of the quark masses.
The relation in Eq.~\ref{eq:Vus} is not scale invariant, 
because of the scale evolution of $V_{us}$,
so may only hold at
specific scale, such as the unification scale.


\begin{figure}
\centering
\includegraphics[%
  scale=0.4, 
  angle=0,keepaspectratio=true,
  origin=t,clip]{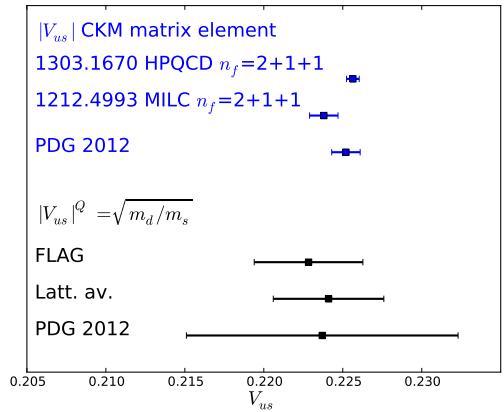}
\caption{Testing $V_{us} \sim \sqrt{\frac{m_d}{m_s}}$}
\label{fg:VusTest}
\end{figure}

In figure~\ref{fg:VusTest} I compare the 
right and left hand sides of 
Eq.~\ref{eq:Vus} with recent results
for quark masses~\cite{Colangelo:2010et,Laiho:2009eu,Beringer:1900zz}
and the $V_{us}$ CKM matrix 
element~\cite{Colangelo:2010et,Bazavov:2013cp,Dowdall:2013rya}.
The errors on the quark masses must be further 
reduced to test the relation in~\ref{eq:Vus}.
The error $m_u/m_d$ needs to be of the order of the 
preliminary result quoted by the MILC 
collaboration~\cite{Bazavov:2012dg} 
to produce an error with a similar size to that of 
$V_{us}$ in equation~\ref{eq:Vus}.

\section{Anthropic constraints on the quark masees}

If you don't like even trying to use textures 
to ``explain'' the 
CKM matrix and masses of the quarks, there is
a much worse possibility. Although the LHC did
find the Higgs boson, it has so far not found any
evidence for BSM particles,
hence it could be that the 
standard model of particle physics is all there is
and we will never understand the values
of the quark masses.

At the lattice 2002 conference Wilczek 
briefly reviewed~\cite{Wilczek:2002wi}
various anthropic principles and suggested
that the lattice QCD community investigate them.
The subject of anthropics comes with a lot of philosophical baggage 
(such as those associated with the current multi-verse mania)
that I will just ignore. However questions such as 
by how much can the masses of the light 
quarks be varied before deuterium is unbound is a well
defined (and hence scientific) question.
Lattice QCD can map out the quark mass dependence of
physical quantities, so should be able to contribute.
Here I will briefly review what has been done for
constraints on the sum of the light quark masses
($m_{ud}$).
The are also constraints on the difference of the 
mass of the up and down quarks
which I do not discuss~\cite{Damour:2007uv}
(see Ref.~\cite{Borsanyi:2013lga} for a recent lattice calculation.)

The properties of the lighter nuclei can be obtained
from solving an effective field theory or a model,
of protons and neutrons 
interacting~\cite{Lee:2008fa,Savage:2011xk,Epelbaum:2013tta}.
The parameters of the effective field theory
can be extracted from experiment or lattice
QCD calculations. Examples of the parameters of the effective field
theory are
the nucleon axial charge ($g_A$) and nucleon-nucleon
scattering lengths.
The basic idea is that the effective field theory
computes the binding energy of nuclei in terms
of hadronic degrees of freedom. The 
dependence of the nucleon, pion and other
quantities on the quark masses is used to
compute the sensitivity of the nuclei binding
energies on the quark masses.

For example Donoghue and Damour~\cite{Damour:2007uv}
found the constraint
$$
\frac{m_{ud}}{ (m_{ud} )_{\mbox{phys}}   }  <  1.64
$$
at 95\% confidence level, from the existence of 
nuclear binding. 
Jaffe et al.~\cite{Jaffe:2008gd} have also investigated
the effect of varying the quark masses on 
producing the
stable nuclei required to make organic
chemistry possible.

Epelbaum et al.~\cite{Epelbaum:2012iu}
have studied the light quark mass dependence
of the binding energy of the Hoyle state $^{12}C$. The production
of carbon and oxygen in red giant stars requires this state to exist.
From their calculation they find that the production of carbon and
oxygen via the Hoyle state is stable for a 2\% change in the 
light quark masses.

There are also tight constraints on the quark masses form 
from Big Bang 
nucleosynthesis~\cite{Bedaque:2010hr,Berengut:2013nh}.
For example, Berengut et al.~\cite{Berengut:2013nh}
find $\frac{\delta m_{ud}}{ m_{ud} } < 0.02 \pm 0.04 $
for nucleosynthesis.

\subsection{Condensates from lattice QCD}

Lattice QCD calculations also provide 
values for non-perturbative quantities
which are useful input to sum rule calculations,
and other formalisms which are used to solve QCD.
There have been a large number of lattice QCD calculations
of the chiral condensate. See Cichy et al.~\cite{Cichy:2013gja}
for a recent calculation of the chiral condensate and a summary
of the results of other calculations. 
The gluon condensate $\langle \frac{\alpha}{\pi}GG \rangle$ has recently
been estimated from the plaquette~\cite{Horsley:2012ra}
using lattice QCD.

In the last year the ratio of strange to light condensates has been
computed using lattice QCD~\cite{McNeile:2012xh}. The key problem
is that there is a $1/a^2$ divergence in the 
condensate with massive quarks,
which must be subtracted off.
The ratio of the strange to light quark condensate was 
found to be 1.08(16)~\cite{McNeile:2012xh}.

\section{Conclusions}

Lattice QCD calculations have recently produced accurate 
results for quark masses and the strong coupling with
errors at the \% level. These results required
lattice QCD calculations with 
controlled continuum extrapolations,
pion masses below
300 MeV (and even in some cases below the physical
pion mass) and the renormalisation
of the quark masses beyond one loop level.

It is important to
have additional results for quark masses and the QCD
coupling from other discritization of the QCD action,
and in particular results from lattice QCD calculations which 
include the dynamics of the
charm quark in the sea (there are at least two
in progress).

This work is supported by SFB-TR 55.


\end{document}